\documentclass[%
	preprint,%
	12pt,%
	number,%
	twocolumn,
	3p,
	centertitle,%
]{elsarticle}

\biboptions{numbers,super,sort&compress}

\usepackage[utf8]{inputenc}

\usepackage{amssymb}
\usepackage{amsmath}
\usepackage{xfrac}
\newcommand*{\dd}[2][]{\mathop{}\!\mathrm{d}^{#1}\!#2}
\newcommand{\eqncomma}{\ensuremath{~,}}
\newcommand{\eqndot}{\ensuremath{~.}}

\usepackage[single=true]{acro}
\DeclareAcronym{RT}{short=RT,long=radiotherapy}
\DeclareAcronym{TMZ}{short=TMZ,long=temozolomide}
\DeclareAcronym{TPS}{short=TPS,long={treatment planning system}}
\DeclareAcronym{IMRT}{short=IMRT,long={intensity modulated radiotherapy}}
\DeclareAcronym{VMAT}{short=VMAT,long={volumetric modulated arc therapy}}
\DeclareAcronym{3DCRT}{short={3D-CRT},long={3D-conformal radiotherapy}}
\DeclareAcronym{DCA}{short=DCA,long={dynamic conformal arc}}
\DeclareAcronym{TMPRT}{short=TMPRT,long={temporally modulated pulsed radiotherapy}}
\DeclareAcronym{MU}{short=MU,long={monitor unit}}
\DeclareAcronym{HI}{short=HI,long={homogeneity index}}
\DeclareAcronym{CI}{short=CI,long={conformality index}}
\DeclareAcronym{TV}{short=TV,long={target volume}}
\DeclareAcronym{PIV}{short=PIV,long={prescription isodose volume}}
\DeclareAcronym{OAR}{short=OAR,long={organ at risk},plural-form={organs at risk}}
\DeclareAcronym{EPID}{short=EPID,long={electronic portal imaging device}}
\DeclareAcronym{MLC}{short=MLC,long={multi leaf collimator}}
\DeclareAcronym{GBM}{short=GBM,long={glioblastoma}}
\DeclareAcronym{LINAC}{short=LINAC,long={linear accelerator}}

\usepackage{caption}
\usepackage{multirow}
\usepackage{booktabs}
\usepackage{siunitx}
\newcommand{\mypc}[1]{~(\qty{#1}{\percent})}

\usepackage{lineno}

\usepackage{hyperref}
\hypersetup{%
	colorlinks=true,
	linkcolor=purple,
	menucolor=purple,
	citecolor=olive,
}
\journal{Practical Radiation Oncology}

\begin{document}
\modulolinenumbers[5]
	
\begin{frontmatter}
	
	
	
	\title{A primer on treatment planning aspects for temporally modulated pulsed radiation therapy}
	
	\author[mmc,einstein]{Christian Velten}
	\author[wisc]{Adam Bayliss}
	\author[washu]{Jiayi Huang}
	\author[mmc,einstein]{Wolfgang A. Tomé\corref{cor1}}
	\ead{wtome@montefiore.org}
	
	\cortext[cor1]{Corresponding author}
	
	\affiliation[mmc]{
		organization={Montefiore Medical Center, Department of Radiation Oncology},
		city={Bronx},
		state={NY},
		country={USA}
	}
	\affiliation[einstein]{
		organization={Albert Einstein College of Medicine, Institute for Onco-Physics},
		city={Bronx},
		state={NY},
		country={USA}
	}
	\affiliation[wisc]{
		organization={University of Wisconsin, Department of Human Oncology},
		city={Madison},
		state={WI},
		country={USA}
	}
	\affiliation[washu]{
		organization={Washington University, Department of Radiation Oncology},
		city={Saint Louis},
		state={MO},
		country={USA}
	}
	
	\begin{abstract}
		\Ac{TMPRT} delivers conventional fraction doses of radiation using temporally separated pulses of low doses (\qty{<30}{\centi\gray}) yielding fraction-effective dose rates of around \qty{6.7}{\centi\gray\per\minute} with the goal to exploit tumor radiation hypersensitivity, which was observed in both, preclinical models and in human clinical trials.
		To facilitate \ac{TMPRT}, \ac{VMAT} and \acs{3DCRT} planning techniques were developed following the guidelines of the proposed NRG CC\nobreakdashes-017 trial. Plans were evaluated with respect to \acs{HI}, conformality, and adherence to dose constraints. Deliverability of plans was assessed using in-phantom measurements for absorbed dose accuracy at low dose rates and using \acs{EPID} for isodose verification.
		For \ac{VMAT} only single arc plans were found to be acceptable due to otherwise unacceptably heterogeneous field doses, while for \acl{DCA} machine limtations on the number of monitor units per degree require the use of partial arcs for each pulse.
		Delivery of plans at low dose rates (\qty{\leq 100}{MU\per\minute}) was accurate with high $\Gamma$ pass rates on modern \acsp{LINAC} and moderate pass rates on legacy \acsp{LINAC}, in line with their general performance.
		Generally, \acs{VMAT} is preferred to achieve optimal homogeneity, conformality, and \acs{OAR} sparing, while the use of \acs{3DCRT} can increase the availability of \ac{TMPRT} for more patients and clinics.
		
	\end{abstract}
%
%
%
\end{frontmatter}

\renewcommand{\thefootnote}{\fnsymbol{footnote}}

\acresetall
\section{Introduction}\label{sec:introduction}

Over the past decades, radiobiologists and radiation oncologists alike have observed that certain tumors are particularly sensitive to \ac{RT} delivered with low-dose rates of \qty{<1.67}{\centi\gray\per\minute},\cite{Brenner.1991,Pierquin.2001} which supported the use of low-dose rate brachytherapy.
This effect was also demonstrated in preclinical human \ac{GBM} models where increased G2/M arrest was observed with continuous dose rates of \qty{0.67}{\centi\gray\per\minute}.\cite{Schultz.1990,Schultz.1995}
Additionally, investigators at the Gray Laboratory demonstrated that most tumor cells, including \ac{GBM}, exhibit hyper-sensitivity to low radiation doses \qty{< 30}{\centi\gray} due to a lack of ATM pathway activation for DNA repair.\cite{Joiner.2001,Short.2001,Krueger.2007,Marples.2008}
These two effects combined have given rise to a \ac{RT} technique often referred to as pulsed low-dose rate, pulsed reduced-dose rate, or \ac{TMPRT}, where a conventional fraction dose of \qtyrange{180}{300}{\centi\gray} is delivered in pulses of approximately \qty{20}{\centi\gray}, below the hyper-sensitivity threshold of around \qty{30}{\centi\gray} and with pulses sufficiently separated in time to yield low effective-dose rates of approximately \qty{6.67}{\centi\gray\per\minute}.\cite{Tome.2007}
Since the instantaneous dose rate of \ac{RT} delivery is not changed but rather temporally modulated, \ac{TMPRT} may be the more accurate and preferred name for this technique.
The efficacy of this approach has been demonstrated in preclinical tumor models, including \ac{GBM}.\cite{Dilworth.2013,Lee.2013,Meyer.2016}

In addition to preclinical investigation, \ac{TMPRT} has been succesfully used at multiple institutions for reirradiating recurrent cranial and extra-cranial tumors.\cite{Adkison.2011,Bovi.2020,Burr.2020,Burr.2020g6,Lee.2018,Mohindra.2013,Murphy.2017,Richards.2009,Witt.2019,Yan.2018} 
A prospective, single-arm phase II study has shown promising clinical efficacy of the combination of TMPRT and \ac{TMZ} on survival, neurocognitive function, and quality of life in adult patients with newly diagnosed \ac{GBM}.\cite{Almahariq.2020} Another prospective, single-arm phase II study also evaluated \ac{TMPRT} and \ac{TMZ} in adult patients with newly diagnosed \ac{GBM} and recently completed accrual (\href{https://clinicaltrials.gov/study/NCT04747145}{NCT04747145}). There are two other prospective single-arm phase II studies examining \ac{TMPRT} with or without concurrent bevacizumab for reirradiating recurrent high-grade glioma (\href{https://clinicaltrials.gov/study/NCT01743950}{NCT01743950}) or IDH-mutant glioma (\href{https://clinicaltrials.gov/study/NCT05393258}{NCT05393258}).
A recently published evidence-based practice guideline by the American Society for Radiation Oncology (ASTRO) endorsed \ac{TMPRT} as an appropriate \ac{RT} technique for irradiating or reirradiating \ac{GBM}.\cite{Yeboa.2025}

This work is intended as a primer on the treatment planning aspects for \ac{TMPRT} to facilitate its use, specifically as part of the proposed NRG CC-017 trial, a randomized phase III study that compares \ac{TMPRT} to standard \ac{RT} in adult patients with newly diagnosed MGMT-unmethylated \ac{GBM}. \ac{3DCRT} and intensity-modulated planning techniques were evaluated and described as recipes, complete with plan evaluation metrics and measurement-based deliverability verification of these low-dose and low-dose rate pulses on various \acp{LINAC}.


\section{Methods}

Throughout this manuscript absorbed radiation doses are given in centigray (\qty{1}{\centi\gray} equal to \qty{0.01}{\gray}) in keeping with NRG conventions and for consistency with the proposed NRG CC-017 trial.

In \ac{TMPRT} the fraction dose $D_{\mathrm{fx}}$ is delivered in discrete radiation pulses separated in time, resulting in a low fraction-effective dose rate,
\begin{equation*}
	\langle\dot{D}\rangle_{\mathrm{fx}} = 
	\dfrac{D_{\mathrm{fx}}} {t_{\mathrm{b},1}-t_{\mathrm{b},n}+\Delta t} \eqncomma
\end{equation*}
with the beginning of first and last pulses, $t_{\mathrm{b},1}$ and $t_{\mathrm{b},n}$, and the nominal time between pulses, $\Delta t$. For example, a dose of \qty{200}{\centi\gray} delivered in ten pulses, with each pulse start separated by $\Delta t$ of \qty{3}{\minute} results in a dose rate of \qty{6.7}{\centi\gray\per\minute}.
This formula can also be used to calculate the minimum treatment session time required for each \ac{TMPRT} fraction, which is approximately \qty{30}{\minute} in the example above, excluding patient setup and necessary imaging.
For completeness, one can also consider the immediate effective dose rate which is the $\Delta t$-moving average over the delivered dose rate and which should similarly be below the low-dose rate threshold,
\begin{equation*}
	\langle\dot{D}\rangle(t) = \dfrac{1}{\Delta t}\int\limits_{t-\Delta t}^{t}  \dot{D}\left(t^\prime\right) \dd{t^\prime} \eqndot
\end{equation*}


\subsection{Planning techniques}

Planning technique for \ac{TMPRT} were developed as recipes using the Varian Eclipse 16.1 \ac{TPS} for \ac{3DCRT}, simple \& \ac{DCA}, and \ac{VMAT} such that delivery of one field corresponds to one pulse. Deliverability and dose homogeneity over the delivered field in each pulse were primary goals in addition to the evaluation metrics described next. These recipes were validated using the RayStation 2024 (RS 25).

\subsection{Evaluation metrics}

\ac{VMAT} plans were generated according to the planning techniques for three unique \acp{LINAC}: a Varian Trilogy with Millenium\nobreakdashes-120 \ac{MLC}, TrueBeam Edge with HD\nobreakdashes-120 \ac{MLC}, and Halcyon with its layered SX2 \ac{MLC}. For \ac{3DCRT} plans were only generated for the TrueBeam Edge as no difference in machine performance is expected for this technique.
Plans were evaluated using target coverage, homogeneity, and conformity metrics in addition to any applicable \ac{OAR} constraints. 
Here, the target \ac{HI} is following the definition in ICRU 83,\cite{ICRU.83}
\begin{equation}\label{eqn:HI}
	\mathit{HI} = \dfrac{D_{2\,\%}-D_{98\,\%}}{D_{50\,\%}} \eqncomma
\end{equation}
while conformity is assessed using the Paddick \acl{CI} and its two components, the \textsl{modified PITV ratio} ($\mathit{CI_{TV}}$) and the \textsl{overtreatment ratio} ($\mathit{CI_{PIV}}$):\cite{Paddick.2000}
\begin{equation}\label{eqn:CI}
	\mathit{CI}_{\mathrm{Paddick}} = 
		\underbrace{\dfrac{\mathit{TV}\cap\mathit{PIV}}{\mathit{TV}}}_{\mathit{CI_{TV}}}
		\cdot
		\underbrace{\dfrac{\mathit{TV}\cap\mathit{PIV}}{\mathit{PIV}}}_{\mathit{CI_{PIV}}}
			\eqndot
\end{equation}

The metrics and criteria used here and proposed for the clinical implementation of \ac{TMPRT} are summarized in Table~\ref{tab:criteria}. Target and \ac{OAR} constraints closely follow the proposed NRG\nobreakdashes-CC017 trial with absolute dose constraints given in the appendix (Tab.~\ref{tab:nrg_criteria}).

\begin{table}[tbh]
	\centering
	\small
	\caption{Suggested target metrics and criteria for implementation of \ac{TMPRT}. Percentages were calculated to closely follow the NRG\nobreakdashes-CC017 trial.}\label{tab:criteria}
	\begin{tabular}{
			lll
		}
		\toprule
			\multirow{2}{*}{\footnotesize\textbf{Metric}}
			 & {\multirow{2}{*}{\footnotesize\textbf{Constraint}}}
			 &  \footnotesize\textbf{Acceptable} \tabularnewline
			 & & \footnotesize\textbf{Variation} \tabularnewline
		\midrule
			$D_{\qty{95}{\percent}}$ & \qty{\geq 98.75}{\percent} & \qty{\geq 95}{\percent} \tabularnewline
			$D_{\qty{10}{\percent}}$ & \qty{\leq 105}{\percent} & \qty{\leq 108.3}{\percent} \tabularnewline
			$D_{\qty{0.03}{\centi\meter\cubed}}$ & \qty{\leq 106.7}{\percent} & \qty{\leq 110}{\percent} \tabularnewline
		\bottomrule
	\end{tabular}
\end{table}

\subsection{Plan delivery measurements}

Absolute dose verification was performed by performing in\nobreakdashes-phantom ionization chamber measurements using a PTW N31010 on a Varian TrueBeam Edge.
A charge-to-dose conversion factor was calculated by using a \qty[parse-numbers=false]{10 \times 10}{\centi\meter\squared} reference field irradiation with \qty{100}{MU}, which also corrects for daily output variations.
The absorbed dose was measured at least five times for each distinct pulse.
Measurement-based isodose verifications were performed for \ac{VMAT} using \ac{EPID} on three linear accelerators: Varian Trilogy, TrueBeam Edge, and Halcyon. Measurements on the Trilogy unit were performed for portal dosimetry plans generated with \qtylist{100;600}{MU\per\minute} as set dose rates.

\section{Results}\label{sec:results}
\subsection{Planning recipes}
The following recipes were developed for both fractionation regimen in NRG\nobreakdashes-CC017, \qtyrange{4600}{6000}{\centi\gray} in \qty{200}{\centi\gray} fractions (conventional) and \qty{4000}{\centi\gray} in \num{15} fractions (hypofractionation). For any other regimen one may follow the same overall approach.
Field dose rates should be set to the larger of \qty{100}{MU\per\minute} or the lowest  for the machine. Field normalization for dose calculation should be set to \qty{100}{\percent} at isocenter when available in the \ac{TPS}.

\subsubsection{VMAT with single arc}
Enable \ac{TPS} features to increase convergence requirements to improve plan quality for this single arc configuration (e.g., for Eclipse set \textit{convergence mode: "Extended"} and \textit{restart from MR level: MR3}). While the collimator rotation may be chosen freely, alignment more perpendicular to the arc plane will allow for improved modulation, especially for targets with \ac{OAR} overlap in the inferior direction.

\begin{figure*}[tbh]
	\centering
	\includegraphics[keepaspectratio,width=0.8\textwidth]{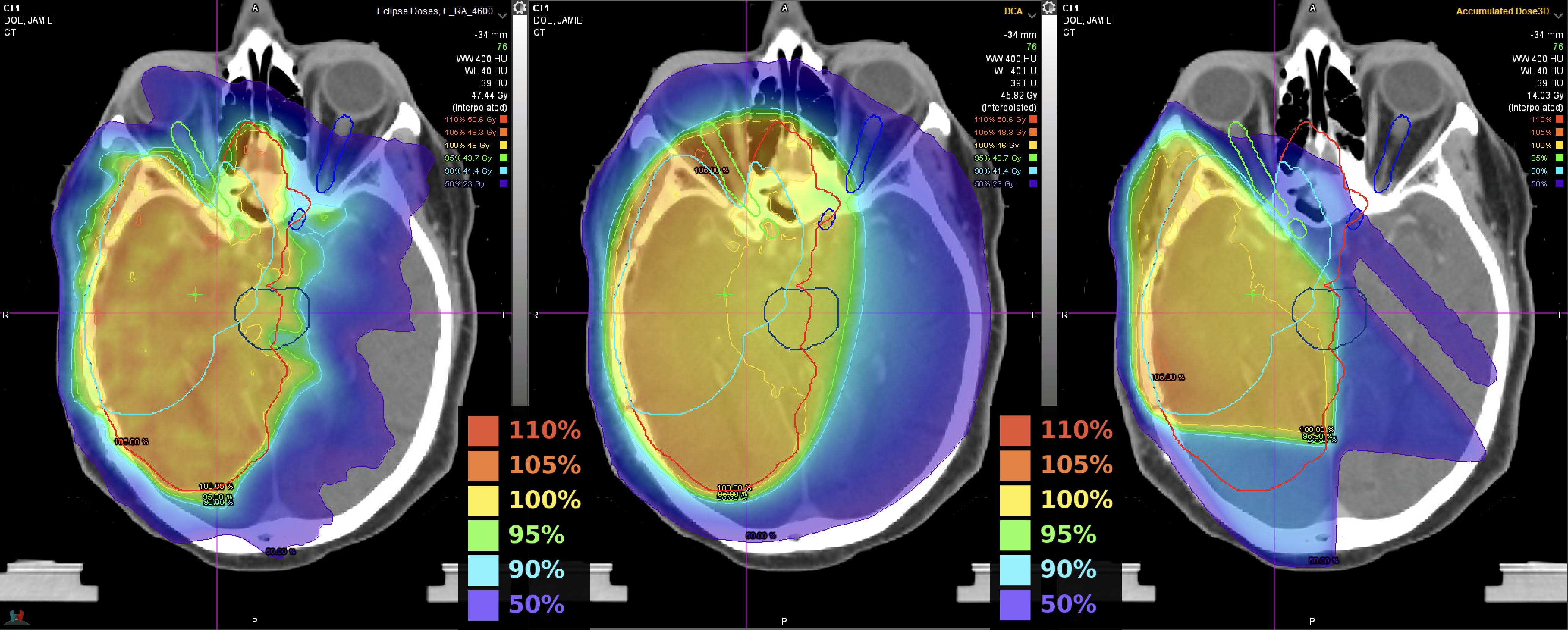}
	\includegraphics[keepaspectratio,width=0.8\textwidth]{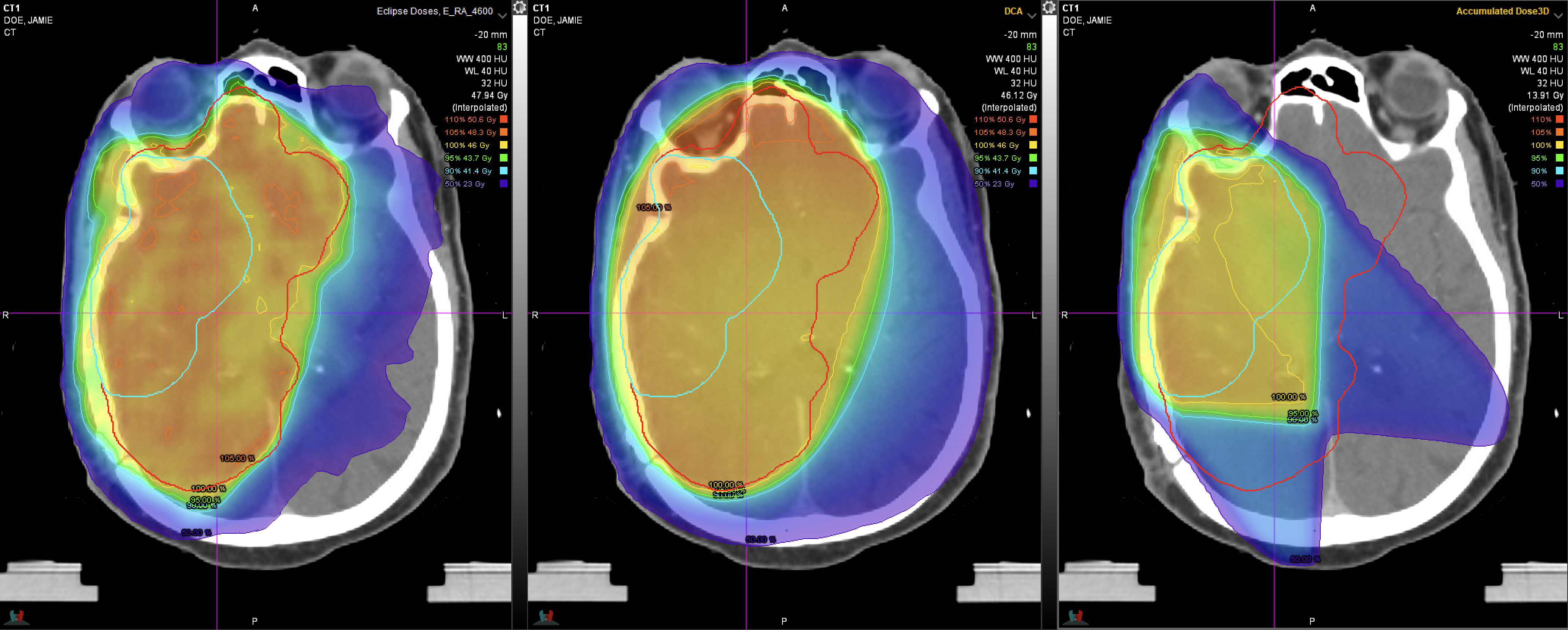}
	\caption{Example dose distributions of single-arc \ac{VMAT} (left) and \ac{DCA} (center) plans to \qty{4600}{\centi\gray}, and a static \ac{3DCRT} (right) plan to \qty{1400}{\centi\gray}. For each two separate axial planes are shown in rows. Structures delineated in red and light blue are PTV4600 and PTV6000, respectively. Colorwash levels are chosen as \qtylist{110;105;100;95;90;50}{\percent} of the respective prescription doses; yellow corresonds to \qty{100}{\percent}.}\label{fig:dosepanels}
\end{figure*}

For conventional fractionation, create a plan with \qty{20}{\centi\gray\per fx} and \num{230} fractions (\num{70} for boost) for a total dose of \qty{4600}{\centi\gray} (\qty{1400}{\centi\gray}) using a single full arc with a control point spacing not exceeding \qty{2}{\degree} (or whichever the minimum is for the \ac{TPS} and treatment machine). Optimization of the plan may be done as usual with the exception that the number of arcs must stay fixed at one. After plan normalization the single arc is copied for a total of ten arcs with the arc direction reversed on alternating arcs (Tab.~\ref{tab:example_beamlist}). Subsequently, the plan is recalculated with fixed \acp{MU} for a prescription of \qty{200}{\centi\gray\per fx} and \num{23} fractions (\num{7} for boost). Each arc within the \qty{200}{\centi\gray\per fx} plan is then delivering one pulse of \qty{20}{\centi\gray}.
The resulting dose distribution for an example single arc \ac{VMAT} plan is shown in the left column of Figure~\ref{fig:dosepanels}.

For hypofractionation, create a plan with \qty{20.5}{\centi\gray\per fx} and \num{195} fractions for a total dose of \qty{3997.5}{\centi\gray} and follow the same procedure except for duplicating the arcs for a total of \num{13} and recalculating with fixed \acp{MU} for a prescription of \qty{4000}{\centi\gray} in \num{15} fractions. This yields a plan where each arc is delivering a pulse of \qty{20.5}{\centi\gray}. Where it is not possible to enter fractional values of centigray, normalization of the final plan should achieve appropriate dosing.

During validation of this recipe with RayStation it was discovered that in this \ac{TPS} the number of fractions is limited to fewer than needed to follow the above recipe. Thus, for RayStation one should create a single arc plan prescribing \qty{20}{\centi\gray} in \num{1} fraction with optimization objectives scaled by \num{0.1}; for hyprofractionation the prescription dose should be \qty{20.5}{\centi\gray} and objectives scaled by \num{0.0769}. After optimization and review, a new plan is generated with the appropriate number of fractions for the overall prescription, e.g., \qty{200}{\centi\gray\per fx} and \num{23} fractions. The previously optimized arc is imported (\textit{copy from}) and subsequently duplicated (\textit{copy beam}) as above.

\subsubsection{3D-DCA with two half arcs}
For conventional fractionation, create a plan with \qty{40}{\centi\gray\per fx} and \num{115} fractions (\num{35} for boost) for a total dose of \qty{4600}{\centi\gray} (\qty{1400}{\centi\gray}) using two half arcs with a control point spacing not exceeding \qty{5}{\degree} (or as above) and a weighting of \qty{50}{\percent} each. After dose calculation and normalization, create copies of each arc for a total of ten arcs with the arc direction reversed on alternating arcs and their weights set to \qty{10}{\percent}. The plan is then recalculated with fixed \acp{MU} for a prescription of \qty{200}{\centi\gray\per fx} as above (Tab.~\ref{tab:example_beamlist}).
The resulting dose distribution for an example \ac{DCA} plan is shown in the center column of Figure~\ref{fig:dosepanels}.

For hypofractionation, create a plan with \qty{38}{\centi\gray\per fx} and \num{105} fractions and follow the same steps except for duplicating the arcs for a total of \num{14} and recalculating with fixed \acp{MU} for a prescription of \qty{4000}{\centi\gray} in \num{15} fractions.

\subsubsection{3D with static fields and accessories}
Create a plan with \qty{200}{\centi\gray\per fx} and \num{23} fractions (\num{7} for boost) for conventional fractionation and with \qty{267}{\centi\gray\per fx} and \num{15} fractions for hypofractionation. Place three fields based on target geometry and avoidance of overlapping \acp{OAR}.

Before manipulating the fields, a pulse distribution for the three fields must be chosen. There are \num{10} pulses for conventional fractionation and \num{13} to \num{14} for hypofractionation. The number of pulses assigned to each field determines its weight to be entered e.g., four pulses in conventional fractionation correspond to a weight of \qty{40}{\percent} whereas four out of \num{14} pulses in hypofractionation corresponds to a weight of \num[parse-numbers=false]{4/14}, or \qty{\approx 28.6}{\percent}. Examples of pulse assignment and corresponding field weights are given in Table~\ref{tab:example_pulseassignment}.
After pulse assignment and weighting the fields may be manipulated using accessories like wedges, field-in-field, or electronic compensator. With the latter, care must be taken to keep the relative field weights equal to maintain the same dose delivered in each pulse.

\begin{table}[tbh]
	\centering
	\small
	\caption{Example of pulse assignment for static \ac{3DCRT} with corresponding field weighting. Conventional fractionation uses \num{10} pulses whereas hypofractionation uses \num{14} pulses.}\label{tab:example_pulseassignment}
	\begin{tabular}{c|
			*{3}{S[table-alignment=right,
				round-mode=places,
				round-precision=1,
				table-format=1.0,
				zero-decimal-to-integer,
				input-symbols={()\%},,
				table-space-text-post={\enspace(11\,\%)},
				table-align-text-post=true]}}
		\toprule
		& \multicolumn{3}{c}{\footnotesize\textbf{pulse assignment (weight)}} \tabularnewline
		\footnotesize\textbf{\#pulses} & \multicolumn{1}{c}{\footnotesize\textbf{Field 1}}
		& \multicolumn{1}{c}{\footnotesize\textbf{Field 2}}
		& \multicolumn{1}{c}{\footnotesize\textbf{Field 3}} \tabularnewline
		\midrule
		10 & 4 \mypc{40} & 4 \mypc{40} & 2 \mypc{20} \tabularnewline
		10 & 4 \mypc{40} & 3 \mypc{30} & 3 \mypc{30} \tabularnewline
		10 & 5 \mypc{50} & 3 \mypc{30} & 2 \mypc{20} \tabularnewline
		13 & 5 \mypc{38} & 5 \mypc{38} & 3 \mypc{24} \tabularnewline
		13 & 5 \mypc{38} & 4 \mypc{31} & 4 \mypc{31} \tabularnewline
		14 & 5 \mypc{35.7} & 5 \mypc{35.7} & 4 \mypc{28.6} \tabularnewline
		14 & 6 \mypc{42.8} & 4 \mypc{28.6} & 4 \mypc{28.6} \tabularnewline
		14 & 6 \mypc{42.8} & 5 \mypc{35.7} & 3 \mypc{21.5} \tabularnewline
		\bottomrule
	\end{tabular}
\end{table}

Once an acceptable plan has been generated, each of the fields should be copied according to its pulse assignment. For example, a field with \num[parse-numbers=false]{4/10} pulses will be copied for total of \num{4} identical fields or a field with \num[parse-numbers=false]{5/14} pulses will be copied for a total of \num{5} identical fields (Tab.~\ref{tab:example_beamlist}). The field weights should then be normalized to sum to \qty{100}{\percent}. For conventional fractionation this means that each field will have a weight of \qty{10}{\percent}.
The resulting dose distribution for an example static \ac{3DCRT} boost plan is shown in the right column of Figure~\ref{fig:dosepanels}.

\subsection{Plan evaluation metrics}

Plans were created for conventional fractionation including the initial and boost plans. \Ac{VMAT} was used on three different units whereas for \ac{3DCRT} only one set of plans was created using \ac{DCA} for the initial and static fields for the boost plan. 

Most plans complied with ideal constraints and achieved similar heterogeneity and conformity metrics tabulated in Table~\ref{tab:plan_evaluation}. Where they did not acceptable variations were met for $D_{\qty{10}{\percent}}$ and $D_{\qty{0.03}{\centi\meter\cubed}}$. In order to comply with dose constraints for brainstem, optic chiasm and nerves, the \ac{3DCRT} boost plan did not achieve full coverage as indicated by the low $\mathit{CI_{TV}}$; it did, however, achieve the protocol's variation acceptable for $D_{\qty{95}{\percent}}$.
%
Homogeneities in \ac{VMAT} plans treating the \qtylist{4600;6000}{\centi\gray} volumes were \numlist{\leq 0.8;\leq 0.13}, respectively. In these plans $\mathit{CI}_{\mathrm{Paddick}}$ were \numlist{\geq 0.91;\geq 0.84} with differences driven by variations in $\mathit{CI_{PIV}}$.

\begin{table*}[bth]
	\centering
	\small
	\caption{Evaluation metrics for conventional fractionation plans generated per the recipes. Absorbed doses are given in percent of prescription unless explicitly specified. For \ac{3DCRT}, \ac{DCA} was used for the initial plan and static fields were used for the boost plan. Differences between the stated values for $\mathit{CI}$ and its factors are due to rounding.\\ *met variation acceptable only}\label{tab:plan_evaluation}
	\begin{tabular}{
			ll
			*{3}{S[round-mode=places,round-precision=1,table-format=3.1]}
			*{4}{S[round-mode=places,round-precision=2,table-format=1.2]}
		}
		\toprule
			\multicolumn{1}{c}{\footnotesize $\boldsymbol{D_{\mathrm{Rx}}}$}
			& \multicolumn{1}{c}{\footnotesize\textbf{Plan (unit)}}
			& {\footnotesize $\boldmath D_{\qty{95}{\percent}}$}
			& {\footnotesize $\boldmath D_{\qty{10}{\percent}}$}
			& {\footnotesize $\boldmath D_{\qty[round-precision=2]{0.03}{\centi\meter\cubed}}$}
			& {\footnotesize $\boldsymbol{\mathit{HI}}$}
			& {\footnotesize $\boldsymbol{\mathit{CI}_{\mathrm{Paddick}}}$}
			& {\footnotesize $\boldsymbol{\mathit{CI_{TV}}}$}
			& {\footnotesize $\boldsymbol{\mathit{CI_{PIV}}}$}
			 \tabularnewline
		\midrule
		
			\qty{4600}{\centi\gray} & VMAT (Edge) & 	 100.10 	&	103.9	&	107.1\textsuperscript{*}	&	0.058	&	0.92	&	0.95	&	0.96	\tabularnewline
			\qty{4600}{\centi\gray} & VMAT (Halcyon) & 	 100.00 	& 105.4\textsuperscript{*}	&	109.5\textsuperscript{*}	&	0.079	&	0.91	&	0.95	&	0.96	\tabularnewline
			\qty{4600}{\centi\gray} & VMAT (Trilogy) & 	 100.00 	&	104.5	&	107.9\textsuperscript{*}	&	0.066	&	0.91	&	0.95	&	0.96	\tabularnewline
			\qty{6000}{\centi\gray} & VMAT (Edge) & 	 99.10 	&	103.8	&	106.4	&	0.112	&	0.86	&	0.94	&	0.92	\tabularnewline
			\qty{6000}{\centi\gray} & VMAT (Halcyon) & 	 100.00 	&	106.4\textsuperscript{*}	&	109.4\textsuperscript{*}	&	0.125	&	0.84	&	0.95	&	0.88	\tabularnewline
			\qty{6000}{\centi\gray} & VMAT (Trilogy) & 	 99.80 	&	105.1\textsuperscript{*}	&	108.3\textsuperscript{*}	&	0.119	&	0.84	&	0.95	&	0.88	\tabularnewline
			
			\midrule
			
			\qty{4600}{\centi\gray} & 3D-DCA (Edge) & 	 98.80 	&	105.4\textsuperscript{*}	&	110\textsuperscript{*}	&	0.103	&	0.75	&	0.87	&	0.86	\tabularnewline
			\qty{6000}{\centi\gray} & 3D (Edge) & 	 95.74\textsuperscript{*} &	100.0	&	101.9	&	0.132	&	0.059	&	0.10	&	0.59	\tabularnewline

		\bottomrule
	\end{tabular}
\end{table*}

\subsection{Plan delivery measurements}
\begin{table}[tbh]
	\centering
	\small
	\caption{In-phantom ionization chamber measurements in high dose and low gradient regions of individual fields (pulses) of each plan delivered on a TrueBeam Edge unit. Where no uncertainty is given it was less than the significant digit.}\label{tab:chamber_measurements}
	\begin{tabular}{%
		l
		*{2}{S[%
			table-alignment=left,%
			round-mode=places,%
			round-precision=1,
			table-format=3.1(1)]}
		S[%
			table-alignment=left,%
			round-mode=places,%
			round-precision=1,
			table-format=-1.1,]
		}
		\toprule
			\multicolumn{1}{c}{\footnotesize\textbf{Field}} & 
			\multicolumn{1}{c}{\footnotesize$\boldmath D_{\mathrm{meas}}\,/\,\unit{\centi\gray}$} & 
			\multicolumn{1}{c}{\footnotesize$\boldmath D_{\mathrm{TPS}}\,/\,\unit{\centi\gray}$} & 
			\multicolumn{1}{c}{\footnotesize$\boldmath\Delta_{\mathrm{rel}}\,/\,\unit{\percent}$} \tabularnewline
		\midrule
			\multicolumn{4}{c}{\footnotesize\textit{VMAT single arc}} \tabularnewline
			CW & 21.2(1) & 20.9(2) & 1.4 \tabularnewline
			CCW & 21.1(1) & 20.9(2) & 0.9 \tabularnewline 
		\midrule
			\multicolumn{4}{c}{\footnotesize\textit{3D\nobreakdashes-\ac{DCA} with two half arcs}} \tabularnewline
			Arc 1 & 18.0 & 17.9(1) & 0.6 \tabularnewline
			Arc 2 & 23.2 & 23.0(1) & 0.9 \tabularnewline 
		\midrule
			\multicolumn{4}{c}{\footnotesize\textit{3D static with accessories}} \tabularnewline
			A1 & 24.0 & 23.7(2) &  1.3 \tabularnewline
			A2 & 20.3 & 20.4(2) & -0.5 \tabularnewline 
			A3 & 18.7 & 18.6(2) & 0.5 \tabularnewline 
		\midrule
		\bottomrule
	\end{tabular}
\end{table}

Ionization chamber measurements for each pulse of the plans described above are given in Table~\ref{tab:chamber_measurements}. Deviations from the expected mean dose at the chamber area were within \qty{1.4}{\percent} for all plans and pulses. Measurement variations over at least five replicates per configuration were \qty{1}{\percent} or less.

The average fractions of points in the \ac{EPID} measurements with $\Gamma(\qty{1}{\percent},\qty{1}{\milli\meter}){\;<\;}1$ were \qtylist{91.0(0.1);97.2(0.2);88.9(1.6)}{\percent} for the TrueBeam Edge, Halcyon, and Trilogy units, respectively. Only points above a \qty{10}{\percent} dose threshold were included in the analysis and dose distributions were normalized to maximum dose.
When increasing to a clinically more commonly used dose difference threshold of \qty{2}{\percent}, these increased to \qtylist{96.2(0.3);99.2(0.0);96.8(0.8)}{\percent}.
On the Trilogy unit, the plan created with the highest set dose rate available had \qtylist{28.3(4.0);94.3(1.0)}{\percent} of points with $\Gamma{\;<\;}1$ for dose difference thresholds of \qtylist{1;5}{\percent}, respectively.

\begin{figure*}[bth]
	\centering
	\includegraphics[keepaspectratio,width=6in]{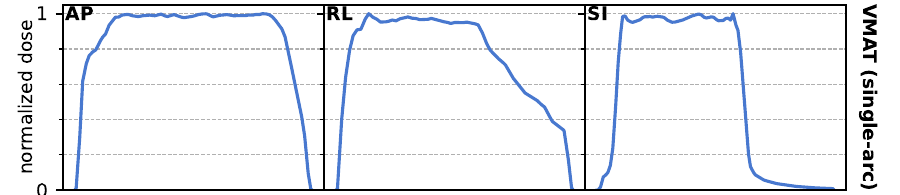}
	\includegraphics[keepaspectratio,width=6in]{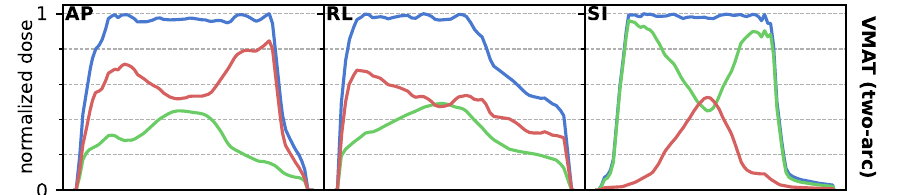}
	\includegraphics[keepaspectratio,width=6in]{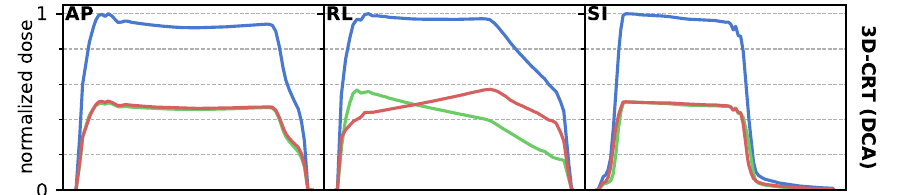}
	\caption{Dose profiles for single- and two-arc \ac{VMAT} (top and middle) and \ac{DCA} (bottom) showing the plan total (blue) and field (red, green) dose profiles in three directions through isocenter. Note that any profile alignment across different plans are coincidental. AP: anterior-posterior; RL: right-left; SI: superior-inferior.}\label{fig:tps_dose_profiles}
\end{figure*}

\section{Discussion}\label{sec:discussion}

\subsection{Planning recipes}

Several beam geometries were initially considered for \ac{VMAT}, specifically two arcs with complementary collimator angles in addition to the single arc setup discussed here. Using complementary collimator angles results in substantially increased degrees of freedom for the optimization, e.g., $N_{\mathrm{dof}}(0,\qty{90}{\degree}){\;\sim\;} N_{\mathrm{dof}}(0)\times N_{\mathrm{dof}}(\qty{90}{\degree})$ compared to $N_{\mathrm{dof}}(0,0){\;\sim\;} 2\,N_{\mathrm{dof}}(0)$. This is typically desirable to simultaneously improve target dose conformality and homogeneity, and critical \ac{OAR} avoidance.
We found, however, that this can lead to substantial differences in field doses of each arc as shown in Figures~\ref{fig:tps_dose_profiles} and~\ref{fig:TwoArcBad}. Furthermore, it is not commonly possible to apply optimization constraints on homogeneity to indvidual fields\footnote{RayStation 2024 and later allows for beam-specific constraints}, nor is a per-field dose evaluation commonly performed by end users in the \ac{TPS}.
This poses an issue for \ac{TMPRT} where each pulse is supposed to deliver a portion of the fraction dose, around \qtyrange{15}{25}{\centi\gray}, to exploit radiation hypersensitivity.
Reviewing the anatomy of several \ac{GBM} cases informed the recommendation to use collimator angles for single arc \ac{VMAT} that result in the \ac{MLC} leaves moving along the superior-inferior direction i.e., around \qty{90}{\degree} (IEC scale).
Users of the RayStation \ac{TPS} (RS 15 or newer) may utilize beam-specific clinical goals or objectives when available. These have the potential to enable multiple arc \ac{TMPRT} optimization, which could be beneficial in cases of substantial overlap between critical \acp{OAR} and the \qty{6000}{\centi\gray} target volume. However, for consistency the NRG CC\nobreakdashes-017 trial allows only identical segment/arc planning for \ac{VMAT}. When using more than a single arc for \ac{TMPRT} treatments outside of that trial, the \ac{HI} and \ac{CI} of each individual segment should be verified and reported.

Splitting the \ac{DCA} delivery into two half arcs was required due to machine limitations on all considered units, specifically the lower limit for delivered monitor units per degree. On the more modern TrueBeam Edge and Halcyon systems this limit is \qty{0.1}{MU\per\degree}, while for older units like the Trilogy it may be higher (\qty{0.3}{MU\per\degree}). Other manufacturers' \acp{LINAC} may have similar limitations, which must be considered individually when commissioning \ac{TMPRT}. While this limit also applies to \ac{VMAT} arcs they typically require more than \qty{36}{MU} to accommodate modulation.
In contrast to the two-arc \ac{VMAT} solution, however, using two \acp{DCA} did not yield substantially different field doses as shown in Figures~\ref{fig:tps_dose_profiles} and~\ref{fig:field_doses_dca}. Profiles along the anterior-posterior and superior-inferior directions were almost identical. Only in the left-right direction a gradient was observed across the field, which is expected since the delivery consists of partial arcs covering the left and right hemispheres. A similar behavior was observed for the static \ac{3DCRT} plan with field doses and profiles shown in Figure~\ref{fig:field_doses_3d}.

\Ac{VMAT} using a single arc presents the overall best treatment planning and delivery technique for \ac{TMPRT} due to its flexibility in achieving conformal and homogeneous dose distributions and simultaneous \ac{OAR} sparing. \Ac{3DCRT} techniques can provide an acceptable fallback solution where insurance or technical limitations (no \ac{VMAT} capability) exist. While helical tomotherapy delivery is excluded in the proposed trial, users of those units may refer to a previous publication for off-trial treatments.\cite{Rasmussen.2010,Rong.2011}

\subsection{Achieved plan metrics}
Despite the constraints of using only one \ac{VMAT} arc high plan quality was achieved for the three diverse \acp{LINAC}, each using a different \ac{MLC} model.
There was no clear trend of either units yielding better plan metrics (Tab.~\ref{tab:plan_evaluation}).

For \ac{3DCRT} a combination of \ac{DCA} and static fields with dynamic wedges and subfields was chosen primarily to demonstrate both planning techniques. The utility of \ac{DCA} strongly depends on the target shape but requires less planning effort to generate a mostly uniform dose distribution compared to static fields. On the other hand, sparing of critical \acp{OAR} like optic structures and the brainstem, which are often adjacent or encompassed in the target margins is often more easily accomplished using static fields with \ac{MLC} manipulation.

While the recommendations presented are based on a single patient's geometry, we reviewed several patients to select the most challenging target geometry with respect to total volume, shape, and overlap with critical \acp{OAR}. In any case, it will be important to consider each patient's individual anatomy in the planning process.

\subsection{Plan delivery measurements}
Agreement between measured point doses and \ac{TPS} values in high dose, low gradient areas showed that treatment delivery at low dose rates of \qty{100}{MU\per\minute} (or lower during \ac{VMAT}) is consistent and per plan. Differences for \ac{VMAT} delivery were higher than for \ac{DCA} but not larger than \qty{1.4}{\percent}, highlighting that \ac{MLC} position uncertainty, dose rate and gantry speed variations are likely to have negligible impact.
Uncertainties in the position and attenuation properties of the used support structures could contribute to the observed differences, which is corroborated by the larger variation being observed for the posterior wedged field (A1) in static \ac{3DCRT}.

Portal dosimetry modes on the more modern TrueBeam and Halcyon units use a single calibration factor across all dose rates. This is possible since their \ac{EPID} response does not vary substantially across the typical range of dose rates in conventional \ac{RT}. This is supported by the high $\Gamma$\nobreakdashes-pass rates for \ac{VMAT} plans with average dose rates of \qty{\approx 100}{MU\per\minute} compared to the typical calibration dose rate of \qtyrange{400}{600}{MU\per\minute}.
In contrast, on older units like the Trilogy \ac{EPID} dosimetry modes uses individual calibration factors for each possible set dose rate due to an increased dose rate dependence. While the choice of set dose rate does not affect plan optimization or delivery, it affects \ac{EPID} dosimetry accuracy. This was demonstranted by the large difference between $\Gamma$\nobreakdashes-pass rates for \qty{600}{MU\per\minute} (\qty{28}{\percent}) and \qty{100}{MU\per\minute} (\qty{94}{\percent}) set dose rates, where in both cases the machine delivered dose rate was fluctuating around \qty{100}{MU\per\minute}. It is thus recommended to choose the set dose rate closest to the expected average.

\section{Conclusions}\label{sec:conclusions}

Treatment planning for \ac{TMPRT} was shown to be feasible with single-arc \ac{VMAT} and \ac{3DCRT}. \Ac{VMAT} is preferred to achieve optimal homogeneity, conformality, and \ac{OAR} sparing. Measurement-based plan verification of low-dose rate field/pulse delivery showed that both modern and legacy \acp{LINAC} are capable to deliver \ac{TMPRT} plans with high fidelity.

\footnotesize
\bibliography{references.bib}

@article{Almahariq.2020, 
year = {2020}, 
title = {{Pulsed radiation therapy for the treatment of newly diagnosed glioblastoma}}, 
author = {Almahariq, Muayad F and Quinn, Thomas J and Arden, Jessica D and Roskos, P T and Wilson, George D and Marples, Brian and Grills, Inga S and Chen, Peter Y and Krauss, Daniel J and Chinnaiyan, Prakash and Dilworth, Joshua T}, 
journal = {Neuro-Oncology}, 
issn = {1522-8517}, 
doi = {10.1093/neuonc/noaa165}, 
pmid = {32658268}, 
pmcid = {PMC7992887}, 
pages = {447--456}, 
number = {3}, 
volume = {23}
}

@article{Schultz.1990, 
year = {1990}, 
title = {{Radioresponse of human astrocytic tumors across grade as a function of acute and chronic irradiation}}, 
author = {Schultz, Christopher T. and Geard, Charles R.}, 
journal = {International Journal of Radiation Oncology*Biology*Physics}, 
issn = {0360-3016}, 
doi = {10.1016/0360-3016(90)90350-s}, 
pmid = {2262364}, 
pages = {1397--1403}, 
number = {6}, 
volume = {19}
}

@article{Brenner.1991, 
year = {1991}, 
title = {{Conditions for the equivalence of continuous to pulsed low dose rate brachytherapy}}, 
author = {Brenner, D.J and Hall, E.J}, 
journal = {International Journal of Radiation Oncology*Biology*Physics}, 
issn = {0360-3016}, 
doi = {10.1016/0360-3016(91)90158-z}, 
pmid = {1993627}, 
pages = {181--190}, 
number = {1}, 
volume = {20}
}

@article{Schultz.1995, 
year = {1995}, 
title = {{Iododeoxyuridine radiosensitization of human glioblastoma cells exposed to acute and chronic gamma irradiation: mechanistic implications and clinical relevance.}}, 
author = {Schultz, C J and Gaffney, D K and Lindstrom, M J and Kinsella, T J}, 
journal = {The cancer journal from Scientific American}, 
issn = {1081-4442}, 
pmid = {9166468}, 
pages = {151--61}, 
number = {2}, 
volume = {1}
}

@article{Paddick.2000, 
year = {2000}, 
title = {{A simple scoring ratio to index the conformity of radiosurgical treatment plans: Technical note}}, 
author = {Paddick, Ian}, 
journal = {Journal of Neurosurgery}, 
issn = {0022-3085}, 
doi = {10.3171/jns.2000.93.supplement\_3.0219}, 
pmid = {11143252}, 
url = {https://thejns.org/view/journals/j-neurosurg/93/supplement\_3/article-p219.xml}, 
pages = {219--222}, 
number = {supplement\_3}, 
volume = {93}
}

@article{Joiner.2001, 
year = {2001}, 
title = {{Low-dose hypersensitivity: current status and possible mechanisms}}, 
author = {Joiner, Michael C and Marples, Brian and Lambin, Philippe and Short, Susan C and Turesson, Ingela}, 
journal = {International Journal of Radiation Oncology*Biology*Physics}, 
issn = {0360-3016}, 
doi = {10.1016/s0360-3016(00)01471-1}, 
pmid = {11173131}, 
pages = {379--389}, 
number = {2}, 
volume = {49}
}

@article{Short.2001, 
year = {2001}, 
title = {{Low-dose hypersensitivity after fractionated low-dose irradiation in vitro}}, 
author = {Short, S C and Kelly, J and Mayes, C R and Woodcock, M and Joiner, M C}, 
journal = {International Journal of Radiation Biology}, 
issn = {0955-3002}, 
doi = {10.1080/09553000110041326}, 
pmid = {11403705}, 
pages = {655--664}, 
number = {6}, 
volume = {77}
}

@article{Pierquin.2001, 
year = {2001}, 
title = {{Curie medal lecture 2000 The optimization of delivered dose in radiotherapy: is it related to low dose rate?}}, 
author = {Pierquin, Bernard}, 
journal = {Radiotherapy and Oncology}, 
issn = {0167-8140}, 
doi = {10.1016/s0167-8140(00)00271-1}, 
pmid = {11165675}, 
pages = {7--9}, 
number = {1}, 
volume = {58}
}

@article{Krueger.2007, 
year = {2007}, 
title = {{Role of Apoptosis in Low-Dose Hyper-radiosensitivity}}, 
author = {Krueger, S. A. and Joiner, M. C. and Weinfeld, M. and Piasentin, E. and Marples, B.}, 
journal = {Radiation Research}, 
issn = {0033-7587}, 
doi = {10.1667/rr0776.1}, 
pmid = {17316076}, 
pages = {260--267}, 
number = {3}, 
volume = {167}
}

@article{Tome.2007, 
year = {2007}, 
title = {{On the possible increase in local tumour control probability for gliomas exhibiting low dose hyper-radiosensitivity using a pulsed schedule}}, 
author = {Tomé, W A and Howard, S P}, 
journal = {The British Journal of Radiology}, 
issn = {0007-1285}, 
doi = {10.1259/bjr/15764945}, 
pmid = {16945935}, 
pages = {32--37}, 
number = {949}, 
volume = {80}
}

@article{Marples.2008, 
year = {2008}, 
title = {{Low-Dose Hyper-Radiosensitivity: Past, Present, and Future}}, 
author = {Marples, Brian and Collis, Spencer J.}, 
journal = {International Journal of Radiation Oncology*Biology*Physics}, 
issn = {0360-3016}, 
doi = {10.1016/j.ijrobp.2007.11.071}, 
pmid = {18374221}, 
pages = {1310--1318}, 
number = {5}, 
volume = {70}
}

@article{Richards.2009, 
year = {2009}, 
title = {{Pulsed reduced dose-rate radiotherapy: a novel locoregional retreatment strategy for breast cancer recurrence in the previously irradiated chest wall, axilla, or supraclavicular region}}, 
author = {Richards, Gregory M. and Tomé, Wolfgang A. and Robins, H. Ian and Stewart, James A. and Welsh, James S. and Mahler, Peter A. and Howard, Steven P.}, 
journal = {Breast Cancer Research and Treatment}, 
issn = {0167-6806}, 
doi = {10.1007/s10549-008-9995-3}, 
pmid = {18389365}, 
pages = {307--313}, 
number = {2}, 
volume = {114}
}

@article{Rasmussen.2010, 
year = {2010}, 
title = {{Reirradiation of Glioblastoma through the Use of a Reduced Dose Rate on a Tomotherapy Unit}}, 
author = {Rasmussen, Karl H. and Hardcastle, Nicholas and Howard, Steven P. and Tomé, Wolfgang A.}, 
journal = {Technology in Cancer Research \& Treatment}, 
issn = {1533-0346}, 
doi = {10.1177/153303461000900409}, 
pmid = {20626205}, 
pmcid = {PMC2906824}, 
pages = {399--406}, 
number = {4}, 
volume = {9}
}

@article{ICRU.83, 
year = {2010}, 
title = {{ICRU Report 83: Prescribing, recording, and reporting photon-beam intensity-modulated radiation therapy (IMRT)}}, 
author = {{International Commission on Radiation Units and Measurements}}
}

@article{Rong.2011, 
year = {2011}, 
title = {{Treatment Planning for Pulsed Reduced Dose-Rate Radiotherapy in Helical Tomotherapy}}, 
author = {Rong, Yi and Paliwal, Bhudatt and Howard, Steven P. and Welsh, James}, 
journal = {International Journal of Radiation Oncology*Biology*Physics}, 
issn = {0360-3016}, 
doi = {10.1016/j.ijrobp.2010.05.055}, 
pmid = {20884127}, 
pages = {934--942}, 
number = {3}, 
volume = {79}
}

@article{Adkison.2011, 
year = {2011}, 
title = {{Reirradiation of Large-Volume Recurrent Glioma With Pulsed Reduced-Dose-Rate Radiotherapy}}, 
author = {Adkison, Jarrod B. and Tomé, Wolfgang and Seo, Songwon and Richards, Gregory M. and Robins, H. Ian and Rassmussen, Karl and Welsh, James S. and Mahler, Peter A. and Howard, Steven P.}, 
journal = {International Journal of Radiation Oncology*Biology*Physics}, 
issn = {0360-3016}, 
doi = {10.1016/j.ijrobp.2009.11.058}, 
pmid = {20472350}, 
pages = {835--841}, 
number = {3}, 
volume = {79}
}

@article{Mohindra.2013, 
year = {2013}, 
title = {{Wide-field pulsed reduced dose rate radiotherapy (PRDR) for recurrent ependymoma in pediatric and young adult patients.}}, 
author = {Mohindra, Pranshu and Robins, H Ian and Tomé, Wolfgang A and Hayes, Lori and Howard, Steven P}, 
journal = {Anticancer research}, 
pmid = {23749916}, 
pages = {2611--8}, 
number = {6}, 
volume = {33}
}

@article{Lee.2013, 
year = {2013}, 
title = {{Pulsed Versus Conventional Radiation Therapy in Combination With Temozolomide in a Murine Orthotopic Model of Glioblastoma Multiforme}}, 
author = {Lee, David Y. and Chunta, John L. and Park, Sean S. and Huang, Jiayi and Martinez, Alvaro A. and Grills, Inga S. and Krueger, Sarah A. and Wilson, George D. and Marples, Brian}, 
journal = {International Journal of Radiation Oncology*Biology*Physics}, 
issn = {0360-3016}, 
doi = {10.1016/j.ijrobp.2013.04.034}, 
pmid = {23845846}, 
pages = {978--985}, 
number = {5}, 
volume = {86}
}

@article{Dilworth.2013, 
year = {2013}, 
title = {{Pulsed low-dose irradiation of orthotopic glioblastoma multiforme (GBM) in a pre-clinical model: Effects on vascularization and tumor control}}, 
author = {Dilworth, Joshua T. and Krueger, Sarah A. and Dabjan, Mohamad and Grills, Inga S. and Torma, John and Wilson, George D. and Marples, Brian}, 
journal = {Radiotherapy and Oncology}, 
issn = {0167-8140}, 
doi = {10.1016/j.radonc.2013.05.022}, 
pmid = {23791366}, 
pages = {149--154}, 
number = {1}, 
volume = {108}
}

@article{Meyer.2016, 
year = {2016}, 
title = {{Pulsed Radiation Therapy With Concurrent Cisplatin Results in Superior Tumor Growth Delay in a Head and Neck Squamous Cell Carcinoma Murine Model}}, 
author = {Meyer, Kurt and Krueger, Sarah A. and Kane, Jonathan L. and Wilson, Thomas G. and Hanna, Alaa and Dabjan, Mohamad and Hege, Katie M. and Wilson, George D. and Grills, Inga and Marples, Brian}, 
journal = {International Journal of Radiation Oncology*Biology*Physics}, 
issn = {0360-3016}, 
doi = {10.1016/j.ijrobp.2016.04.031}, 
pmid = {27511853}, 
pages = {161--169}, 
number = {1}, 
volume = {96}
}

@article{Murphy.2017, 
year = {2017}, 
title = {{Intensity modulated radiation therapy with pulsed reduced dose rate as a reirradiation strategy for recurrent central nervous system tumors: An institutional series and literature review}}, 
author = {Murphy, Erin S. and Rogacki, Kevin and Godley, Andrew and Qi, Peng and Reddy, Chandana A. and Ahluwalia, Manmeet S. and Peereboom, David M. and Stevens, Glen H. and Yu, Jennifer S. and Kotecha, Rupesh and Suh, John H. and Chao, Samuel T.}, 
journal = {Practical Radiation Oncology}, 
issn = {1879-8500}, 
doi = {10.1016/j.prro.2017.04.003}, 
pmid = {28666902}, 
pages = {e391--e399}, 
number = {6}, 
volume = {7}
}

@article{Yan.2018, 
year = {2018}, 
title = {{Use of Pulsed Low–Dose Rate Radiotherapy in Refractory Malignancies}}, 
author = {Yan, Jing and Yang, Ju and Yang, Yang and Ren, Wei and Liu, Juan and Gao, Shanbao and Li, Shuangshuang and Kong, Weiwei and Zhu, Lijing and Yang, Mi and Qian, Xiaoping and Liu, Baorui}, 
journal = {Translational Oncology}, 
issn = {1936-5233}, 
doi = {10.1016/j.tranon.2017.12.004}, 
pmid = {29306203}, 
pmcid = {PMC5756059}, 
pages = {175--181}, 
number = {1}, 
volume = {11}
}

@article{Lee.2018, 
year = {2018}, 
title = {{Local Control and Toxicity of External Beam Reirradiation With a Pulsed Low-dose-rate Technique}}, 
author = {Lee, Charles T. and Dong, Yanqun and Li, Tianyu and Freedman, Samuel and Anaokar, Jordan and Galloway, Thomas J. and Hallman, Mark A. and Weiss, Stephanie E. and Hayes, Shelly B. and Price, Robert A. and Ma, C.M. Charlie and Meyer, Joshua E.}, 
journal = {International Journal of Radiation Oncology*Biology*Physics}, 
issn = {0360-3016}, 
doi = {10.1016/j.ijrobp.2017.12.012}, 
pmid = {29485075}, 
pmcid = {PMC7537409}, 
pages = {959--964}, 
number = {4}, 
volume = {100}
}

@article{Witt.2019, 
year = {2019}, 
title = {{Large volume re-irradiation for recurrent meningioma with pulsed reduced dose rate radiotherapy}}, 
author = {Witt, Jacob S. and Musunuru, Hima B. and Bayliss, R. Adam and Howard, Steven P.}, 
journal = {Journal of Neuro-Oncology}, 
issn = {0167-594X}, 
doi = {10.1007/s11060-018-03011-z}, 
pmid = {30392090}, 
pages = {103--109}, 
number = {1}, 
volume = {141}
}

@article{Bovi.2020, 
year = {2020}, 
title = {{Pulsed Reduced Dose Rate Radiotherapy in Conjunction With Bevacizumab or Bevacizumab Alone in Recurrent High-grade Glioma: Survival Outcomes}}, 
author = {Bovi, Joseph A. and Prah, Melissa A. and Retzlaff, Amber A. and Schmainda, Kathleen M. and Connelly, Jennifer M. and Rand, Scott D. and Marszalkowski, Cathy S. and Mueller, Wade M. and Siker, Malika L. and Schultz, Christopher J.}, 
journal = {International Journal of Radiation Oncology*Biology*Physics}, 
issn = {0360-3016}, 
doi = {10.1016/j.ijrobp.2020.06.020}, 
pmid = {32599030}, 
pmcid = {PMC8655709}, 
pages = {979--986}, 
number = {4}, 
volume = {108}
}

@article{Burr.2020, 
year = {2020}, 
title = {{Pulsed Reduced Dose Rate for Reirradiation of Recurrent Breast Cancer}}, 
author = {Burr, Adam R. and Robins, H. Ian and Bayliss, R. Adam and Howard, Steven P.}, 
journal = {Practical Radiation Oncology}, 
issn = {1879-8500}, 
doi = {10.1016/j.prro.2019.09.004}, 
pmid = {31526900}, 
pages = {e61--e70}, 
number = {2}, 
volume = {10}
}

@article{Burr.2020g6, 
year = {2020}, 
title = {{Outcomes From Whole-Brain Reirradiation Using Pulsed Reduced Dose Rate Radiation Therapy}}, 
author = {Burr, Adam R. and Robins, Henry Ian and Bayliss, Robert Adam and Baschnagel, Andrew M. and Welsh, James S. and Tomé, Wolfgang A. and Howard, Steven P.}, 
journal = {Advances in Radiation Oncology}, 
issn = {2452-1094}, 
doi = {10.1016/j.adro.2020.06.021}, 
pmid = {33083645}, 
pmcid = {PMC7557211}, 
pages = {834--839}, 
number = {5}, 
volume = {5}
}

@article{Yeboa.2025, 
year = {2025}, 
title = {{Radiation Therapy for WHO Grade 4 Adult-Type Diffuse Glioma: An ASTRO Clinical Practice Guideline}}, 
author = {Yeboa, Debra Nana and Braunstein, Steve E. and Cabrera, Alvin and Crago, Kevin and Galanis, Evanthia and Hattab, Eyas M. and Heron, Dwight E. and Huang, Jiayi and Kim, Michelle M. and Kirkpatrick, John P. and Knisely, Jonathan P.S. and McAleer, Mary Frances and McClelland, Shearwood and Milano, Michael T. and Moliterno, Jennifer and Porter, Alyx and Redmond, Kristin J. and Trifiletti, Daniel M. and Tsien, Christina and Venkatesulu, Bhanu Prasad and Vinogradskiy, Yevgeniy and Bradfield, Lisa and Helms, Amanda R. and Bovi, Joseph A.}, 
journal = {Practical Radiation Oncology}, 
issn = {1879-8500}, 
doi = {10.1016/j.prro.2025.05.014}, 
pmid = {40578479}, 
pages = {451--471}, 
number = {5}, 
volume = {15}
}

\clearpage
\onecolumn
\setcounter{table}{0}
\setcounter{figure}{0}
\renewcommand{\thetable}{A\arabic{table}}
\renewcommand{\thefigure}{A\arabic{figure}}
\section*{Appendix}


\begin{table*}[tbhp]
	\footnotesize
	\caption{Absolute dose target and \ac{OAR} constraints given in centigray as proposed by the NRG\nobreakdashes-CC017 trial. *conventional fractionation; \textsuperscript{\#}hypofractionation}\label{tab:nrg_criteria}
	\begin{tabular}{
			l|ll
			*{4}{l}
		}
		\toprule
		&&& \multicolumn{2}{c}{\footnotesize\textit{conventional fractionation}} & \multicolumn{2}{c}{\footnotesize\textit{hypofractionation}} \tabularnewline
		\footnotesize\textbf{Volume} & \footnotesize\textbf{Metric} & \footnotesize\textbf{Unit} & \footnotesize\textbf{Constraint} & \footnotesize\textbf{Accept. Var.} & \footnotesize\textbf{Constraint} & \footnotesize\textbf{Accept. Var.} \tabularnewline
		\midrule
		
		PTV4600\textsuperscript{*} & $D_{\qty{95}{\percent}}$ & \unit{\centi\gray} & \num{\geq 4600} & \num{\geq 4370} & n/a & n/a \tabularnewline
		\midrule
		
		\multirow{3}{*}{PTV6000\textsuperscript{*}} & $D_{\qty{95}{\percent}}$ & \unit{\centi\gray} & \num{\geq 5925} & \num{\geq 5700} & n/a & n/a \tabularnewline
		& $D_{\qty{10}{\percent}}$ & \unit{\centi\gray} & \num{\leq 6300} & \num{\leq 6500} & n/a & n/a \tabularnewline
		& $D_{\qty{0.03}{\centi\meter\cubed}}$ & \unit{\centi\gray} & \num{\leq 6400} & \num{\leq 6600} & n/a & n/a \tabularnewline
		\midrule
		
		\multirow{3}{*}{PTV4000\textsuperscript{\#}} & $D_{\qty{95}{\percent}}$ & \unit{\centi\gray}  & n/a & n/a & \num{\geq 3955} & \num{\geq 3800} \tabularnewline
		& $D_{\qty{10}{\percent}}$ & \unit{\centi\gray} & n/a & n/a & \num{\leq 4200} & \num{\leq 4340} \tabularnewline
		& $D_{\qty{0.03}{\centi\meter\cubed}}$ & \unit{\centi\gray} & n/a & n/a & \num{\leq 4270} & \num{\leq 4400} \tabularnewline
		\midrule
		
		Brainstem & $D_{\qty{0.03}{\centi\meter\cubed}}$ & \unit{\centi\gray} & \num{\leq 5500} & \num{\leq 6000} & \num{\leq 4500} & \num{\leq 4800} \tabularnewline
		Optic structures (PRV) & $D_{\qty{0.03}{\centi\meter\cubed}}$ & \unit{\centi\gray} & \num{\leq 5500} & \num{\leq 6000} & \num{\leq 4000} & \num{\leq 4200} \tabularnewline
		Retinas (lt/rt) & $D_{\qty{0.03}{\centi\meter\cubed}}$ & \unit{\centi\gray} & \num{\leq 4500} & \num{\leq 5000} & \num{\leq 3500} & \num{\leq 3750} \tabularnewline
		Lenses (lt/rt) & $D_{\qty{0.03}{\centi\meter\cubed}}$ & \unit{\centi\gray} & \num{\leq 1000} & ALARA & \num{\leq 900} & ALARA \tabularnewline
		Spinal cord & $D_{\qty{0.03}{\centi\meter\cubed}}$ & \unit{\centi\gray} & \num{\leq 5000} & none & \num{\leq 4200} & none \tabularnewline
		
		\bottomrule
	\end{tabular}
\end{table*}

\clearpage
\onecolumn
\setcounter{table}{0}
\setcounter{figure}{0}
\renewcommand{\thetable}{S\arabic{table}}
\renewcommand{\thefigure}{S\arabic{figure}}
\section*{Supplementary Material}

\begin{table*}[tbh]
	\centering
	\footnotesize
	\caption{Beam parameter list for the example static and \ac{DCA} \ac{3DCRT}, and (single-arc) \ac{VMAT} plans created following the planning recipes.}\label{tab:example_beamlist}
	\setlength{\tabcolsep}{4pt}
	\begin{tabular}{
			l
			c
			S[table-format=1.3]
			l
			S[table-format=3.1]
			l
			S[table-format=+1.1]
			S[table-format=+1.1]
			S[table-format=+1.1]
			S[table-format=+1.1]
			S[table-format=2.0]
		}
		\toprule
		\footnotesize\textbf{Id} & 
		\footnotesize\textbf{MLC} & 
		{\footnotesize\textbf{Weight}} & 
		{\footnotesize\textbf{Gantry}} & 
		{\footnotesize\textbf{Coll.}} & 
		\footnotesize\textbf{Wedge} & 
		{\footnotesize\textbf{X1}} & 
		{\footnotesize\textbf{X2}} & 
		{\footnotesize\textbf{Y1}} & 
		{\footnotesize\textbf{Y2}} & 
		{\footnotesize\textbf{MU}}
		\tabularnewline
		\midrule
		\multicolumn{11}{c}{\textit{static \ac{3DCRT}}} \tabularnewline
		A1 1 & Static & 0.100 & 180.0{E} & 90.0 & EDW30IN & +3.3 & +3.7 & +4.5 & +3.3 & 30 \tabularnewline
		A1 2 & Static & 0.100 & 180.0{E} & 90.0 & EDW30IN & +3.3 & +3.7 & +4.5 & +3.3 & 30 \tabularnewline
		A1 3 & Static & 0.100 & 180.0{E} & 90.0 & EDW30IN & +3.3 & +3.7 & +4.5 & +3.3 & 30 \tabularnewline
		A2 1 & Static & 0.100 & 320.0 & 345.0 & None & +5.4 & +4.6 & +4.0 & +3.6 & 23 \tabularnewline
		A2 2 & Static & 0.100 & 320.0 & 345.0 & None & +5.4 & +4.6 & +4.0 & +3.6 & 23 \tabularnewline
		A2 3 & Static & 0.100 & 320.0 & 345.0 & None & +5.4 & +4.6 & +4.0 & +3.6 & 23 \tabularnewline
		A2 4 & Static & 0.100 & 320.0 & 345.0 & None & +5.4 & +4.6 & +4.0 & +3.6 & 23 \tabularnewline
		A3 1 & Static & 0.100 & 100.0 & 90.0 & EDW25IN & +3.3 & +3.6 & +5.1 & +5.7 & 27 \tabularnewline
		A3 2 & Static & 0.100 & 100.0 & 90.0 & EDW25IN & +3.3 & +3.6 & +5.1 & +5.7 & 27 \tabularnewline
		A3 3 & Static & 0.100 & 100.0 & 90.0 & EDW25IN & +3.3 & +3.6 & +5.1 & +5.7 & 27 \tabularnewline
		
		\midrule
		\multicolumn{11}{c}{\textit{\acf{DCA}}} \tabularnewline
		A1 CW & Arc Dynamic & 0.100 & 181.0 CW 0.0 & 10.0 &&&&& & 23 \tabularnewline
		A2 CW & Arc Dynamic & 0.100 & 0.0 CW 179.0 & 10.0 &&&&& & 26 \tabularnewline
		A3 CCW & Arc Dynamic & 0.100 & 179.0 CCW 0.0 & 10.0 &&&&& & 26 \tabularnewline
		A4 CCW & Arc Dynamic & 0.100 & 0.0 CCW 181.0 & 10.0 &&&&& & 23 \tabularnewline
		A5 CW & Arc Dynamic & 0.100 & 181.0 CW 0.0 & 10.0 &&&&& & 23 \tabularnewline
		A6 CW & Arc Dynamic & 0.100 & 0.0 CW 179.0 & 10.0 &&&&& & 26 \tabularnewline
		A7 CCW & Arc Dynamic & 0.100 & 179.0 CCW 0.0 & 10.0 &&&&& & 26 \tabularnewline
		A8 CCW & Arc Dynamic & 0.100 & 0.0 CCW 181.0 & 10.0 &&&&& & 23 \tabularnewline
		A9 CW & Arc Dynamic & 0.100 & 181.0 CW 0.0 & 10.0 &&&&& & 23 \tabularnewline
		A10 CW & Arc Dynamic & 0.100 & 0.0 CW 179.0 & 10.0 &&&&& & 26 \tabularnewline
		
		\midrule
		\multicolumn{11}{c}{\textit{\acf{VMAT}}} \tabularnewline
		A1 & VMAT & 0.453 & 181.0 CW 179.0 & 70.0 &&&&& & 91 \tabularnewline
		A2 & VMAT & 0.453 & 179.0 CCW 181.0 & 70.0 &&&&& & 91 \tabularnewline
		A3 & VMAT & 0.453 & 181.0 CW 179.0 & 70.0 &&&&& & 91 \tabularnewline
		A4 & VMAT & 0.453 & 179.0 CCW 181.0 & 70.0 &&&&& & 91 \tabularnewline
		A5 & VMAT & 0.453 & 181.0 CW 179.0 & 70.0 &&&&& & 91 \tabularnewline
		A6 & VMAT & 0.453 & 179.0 CCW 181.0 & 70.0 &&&&& & 91 \tabularnewline
		A7 & VMAT & 0.453 & 181.0 CW 179.0 & 70.0 &&&&& & 91 \tabularnewline
		A8 & VMAT & 0.453 & 179.0 CCW 181.0 & 70.0 &&&&& & 91 \tabularnewline
		A9 & VMAT & 0.453 & 181.0 CW 179.0 & 70.0 &&&&& & 91 \tabularnewline
		A10 & VMAT & 0.453 & 179.0 CCW 181.0 & 70.0 &&&&& & 91 \tabularnewline
		
		\bottomrule
	\end{tabular}
\end{table*}

\clearpage

\begin{figure*}[tbh]
	\centering
	\includegraphics[keepaspectratio,width=0.9\textwidth]{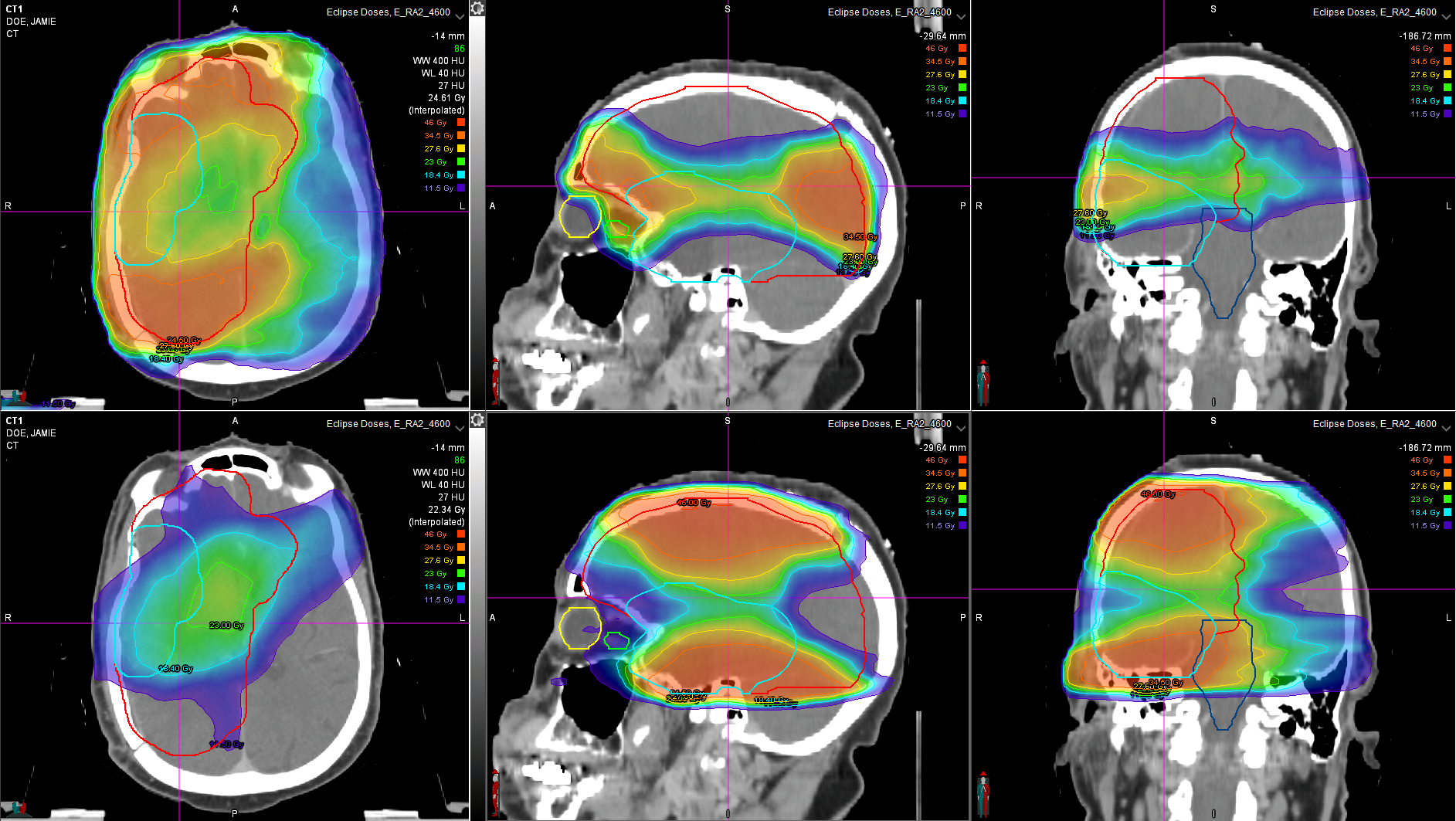}
	\includegraphics[keepaspectratio,width=6in]{figures/dp_vmat}
	\caption{Field doses and profiles for an example two-arc \ac{VMAT} plan. Arcs one and two are shown in the top and bottom row with \acp{HI} of \numlist{1.47;1.60}, respectively. To deliver \qty{4600}{\centi\gray} total, each arc is supposed to contribute \qty{2300}{\centi\gray}; levels in the colorwash are chosen as \qtylist{200;150;120;100;80;50}{\percent} of \qty{2300}{\centi\gray}.}\label{fig:TwoArcBad}
\end{figure*}

\begin{figure*}[tbh]
	\centering
	\includegraphics[keepaspectratio,width=0.9\textwidth]{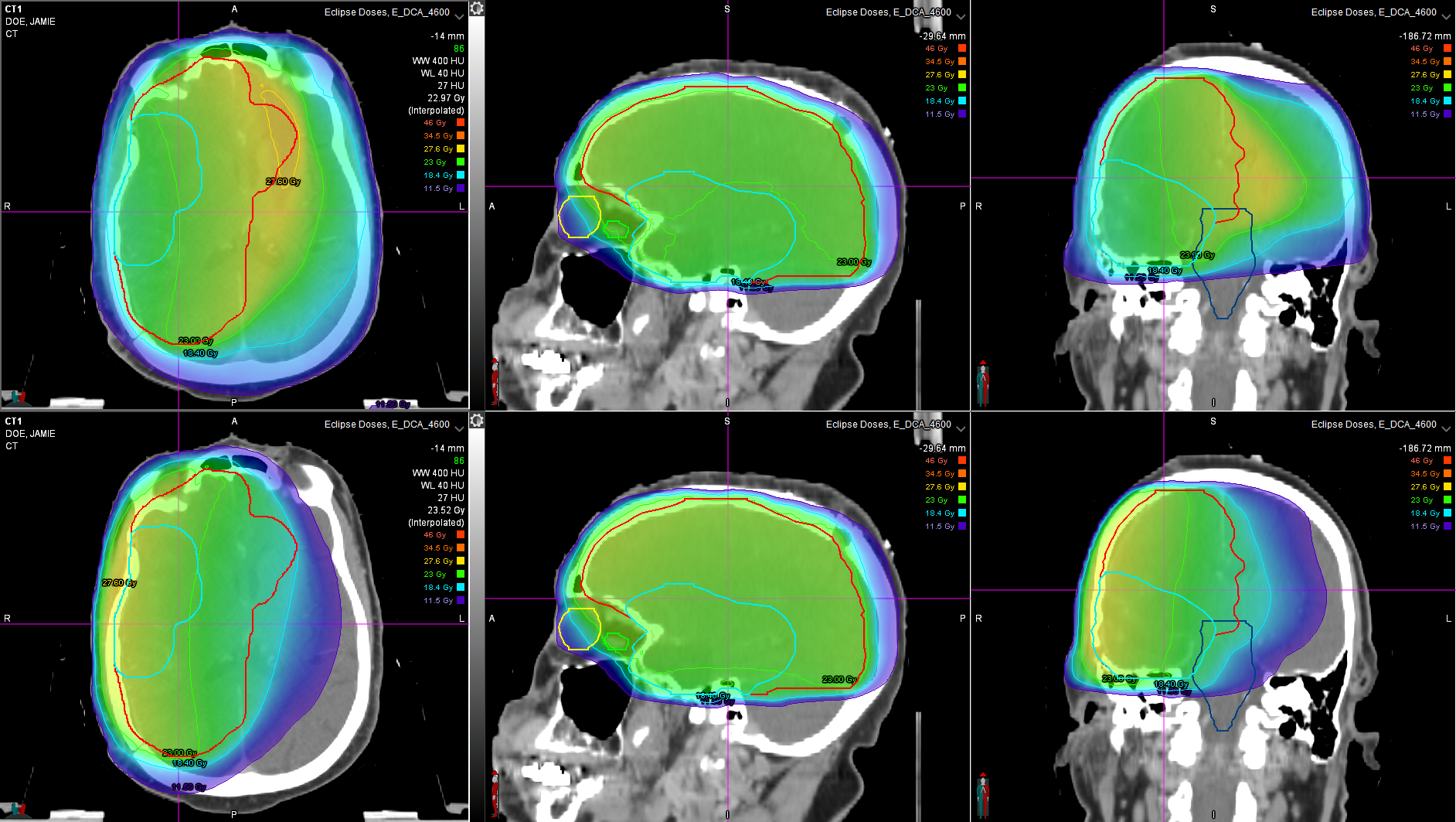}
	\includegraphics[keepaspectratio,width=6in]{figures/dp_dca}
	\caption{Field doses and profiles for an example two-arc \ac{DCA} plan. Arcs one and two are shown in the top and bottom row with \acp{HI} of \numlist{0.28;0.27}, respectively. To deliver \qty{4600}{\centi\gray} total, each arc is supposed to contribute \qty{2300}{\centi\gray}; levels in the colorwash are chosen as \qtylist{200;150;120;100;80;50}{\percent} of \qty{2300}{\centi\gray}.}\label{fig:field_doses_dca}
\end{figure*}

\begin{figure*}[tbh]
	\centering
	\includegraphics[keepaspectratio,width=0.9\textwidth]{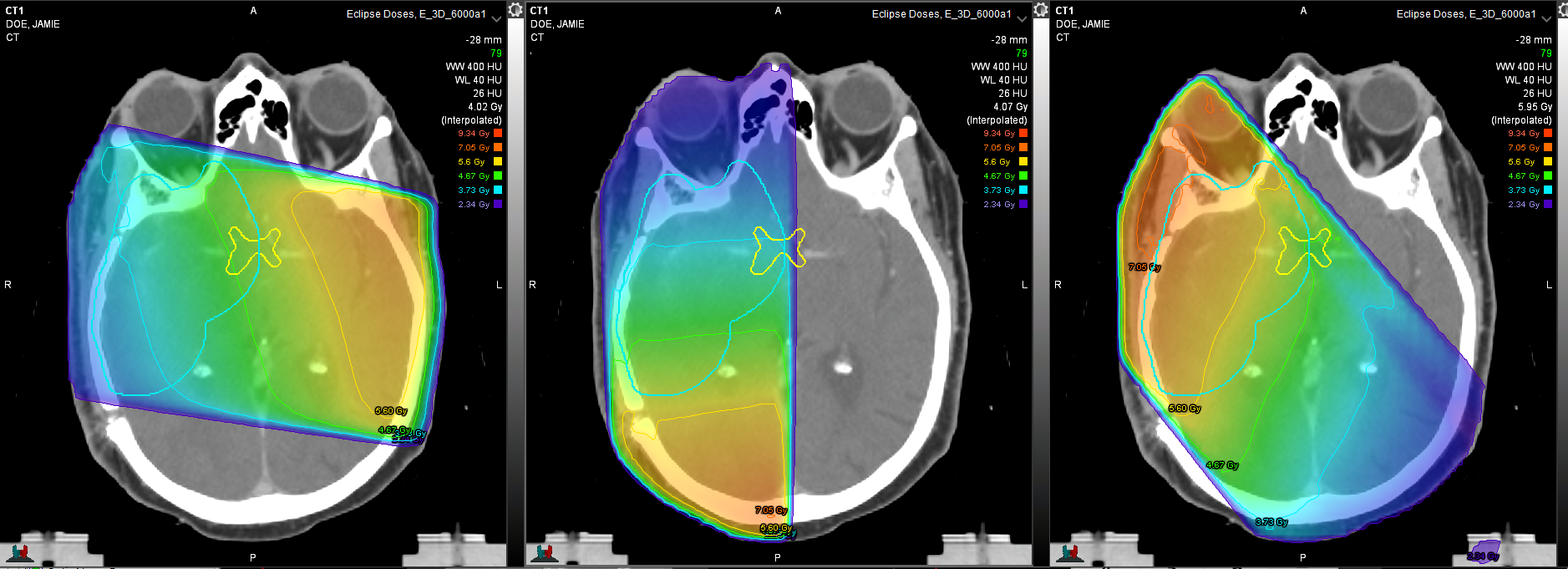}
	\includegraphics[keepaspectratio,width=6in]{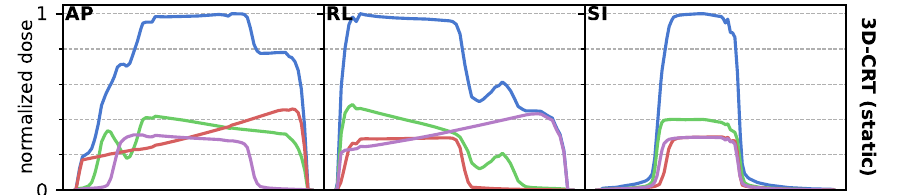}
	\caption{Field doses and profiles through isocenter for an example three-field \ac{3DCRT} boost plan using wedges and \ac{MLC} with \acp{HI} of \numlist{0.38;0.52;0.44}. Levels in the colorwash are chosen as \qtylist{200;150;120;100;80;50}{\percent} of \qty{467}{\centi\gray}.}\label{fig:field_doses_3d}
\end{figure*}


%
%

\end{document}